\documentstyle[12pt]{article}
\newcommand{\be}{\begin{equation}}
\newcommand{\ee}{\end{equation}}
\newcommand{\bea}{\begin{eqnarray}}
\newcommand{\eea}{\end{eqnarray}}
\newcommand{\norsl}{\normalsize\sl}
\newcommand{\norsc}{\normalsize\sc}

\begin{document}

%-------------------- Title page ----------------------------------
\begin{titlepage}

\title{
   Muonium-Antimuonium Conversion\\
    in Models with Dilepton Gauge Bosons}
\bigskip
\author{
\norsc Kuninori HORIKAWA
    and
\norsc  Ken SASAKI\thanks{e-mail address: a121004@c1.ed.ynu.ac.jp}
\\
\norsl  Dept. of Physics, Yokohama National University\\
\norsl  Yokohama 240, JAPAN\\
}

%\date{\today}
\date{}
\maketitle

\begin{abstract}
{\normalsize We examine the magnetic field dependence of the
muonium($\mu^+ e^-$)-antimuonium($\mu^- e^+$) conversion in the models which
accommodate the dilepton gauge bosons. The effective Hamiltonian
for the conversion due to dileptons turns out to be in the
$(V-A)\times(V+A)$ form
and, in consequence, the conversion probability is rather
insensitive to the strength of the magnetic field.
The reduction is less than $20\%$ for
up to $B \approx 300$ G and
$33\%$ even in the large $B$ limit. \\
 \\
PACS number(s): 11.30.Hv, 12.15.Cc, 12.15.Ji, 36.10.Dr}
\end{abstract}

%\begin{picture}(5,2)(-290,-500)
\begin{picture}(5,2)(-290,-550)
\put(2.3,-180){YNU-HEPTh-107}
\put(2.3,-195){March 1995}
\end{picture}

\thispagestyle{empty}
\end{titlepage}
\setcounter{page}{1}
\baselineskip 18pt
%----------------------- Text -----------------------------------

%\numberbysection{\@addtoreset{equation}{section}
%\def\theequation{\arabic{section}.\arabic{equation}}}
%\section{Introduction}
%\smallskip

Muonium, which is a bound state of $\mu^+$ and $e^-$, can be
transformed to antimuonium, a bound state of $\mu^-$ and $e^+$,
if there exists a lepton-number-non-conserving interaction~\cite{Pontecorvo}.
Feinberg and Weinberg~\cite{Feinberg} studied the $M-\overline{M}$ conversion
with a postulated effective Hamiltonian of $(V-A)\times (V-A)$ form.
Later, this process has been studied within the left-right symmetric
models and the models with doubly-charged Higgs
bosons~\cite{Halprin}-\cite{Herczeg}.
In these models, the effective Hamiltonian for the
conversion is expressed either in the $(V-A)\times (V-A)$ form or
in the $(V+A)\times (V+A)$ form. Thus far no $M-\overline{M}$ conversion
has been observed~\cite{Hughes}.

Recently, an interesting class of models which have new $SU(2)_L$-doublet
gauge bosons were proposed as extensions of
the standard model~\cite{Adler}-\cite{Framd}.
In these models each family
of leptons $(l^+, \nu_l, l^-)_L$ transforms as a triplet under the gauge
group $SU(3)$ and the total lepton number defined as $L=L_e+L_{\mu}+L_{\tau}$
is conserved, while the separate lepton number for each family is not.
The new gauge bosons $(X^{\mp}, X^{\mp\mp})$
carry lepton number $L=\pm2$. Hence,
hereafter, we refer to these gauge bosons as dileptons.
The gauge group $SU(3)$ will be, for example, an $SU(3)_l$ in the $SU(15)$
grand unification theory model~\cite{Frama} or an $SU(3)_L$ in the
$SU(3)_C \times SU(3)_L \times U(1)_X$ model~\cite{Framd}.

The phenomenology on dilepton gauge bosons has been extensively studied.
When the doubly-charged dilepton exists, the mixing of
muonium and antimuonium is possible
through the  diagram illustrated in Fig. 1
and thus $M-\overline{M}$ conversion
takes place~\cite{Pal}-\cite{Mimura}. In particular, the effective
Hamiltonian for the mixing turns out to be
in the $(V-A)\times (V+A)$ form.
One of the present authors (K.S.) and Fujii and
Nakamura calculated the probability for the $M-\overline{M}$ conversion in the
models with dileptons and examined the lower mass
bound on the doubly-charged dilepton $X^{\pm \pm}$ in Ref.\cite{Fujii}.
But the analysis
was done in the case of absence of magnetic fields.
In this paper we consider the $M-\overline{M}$ conversion in static
external magnetic fields and study the field dependence of
the conversion probability.

The muonium or antimuonium system in the presence of static
external magnetic field
$\overrightarrow {B}$
is described by the following Hamiltonian,
\be
    {\cal H}_{int}=A\overrightarrow {S_e}\cdot\overrightarrow {S_{\mu}}
            +\mu_{B} g_e \overrightarrow {S_e}\cdot\overrightarrow {B}
          +\mu_{B}\frac{m_e}{m_{\mu}} g_{\mu}
               \overrightarrow {S_{\mu}}\cdot\overrightarrow {B},
\label{HB}
\ee
where $\overrightarrow {S_e}$, $m_e$, $g_{e^-}=-g_{e^+}$ and
$\overrightarrow {S_{\mu}}$, $m_\mu$, $g_{\mu^+}=-g_{\mu^-}$
are spin, mass, the gyromagnetic ratio of electron (or positron) and
$\mu^+$ (or $\mu^-$), respectively, and
$\mu_{B}$ is Bohr magneton.
The first term of Eq.(\ref{HB}) is the source of $1S$
hyperfine splitting of the muonium (or antimuonium) system and
$A=1.846\times 10^{-5}$eV. Taking the magnetic field direction as the
spin-quantization axis, we obtain the muonium energy eigenvalues
as follows~\cite{Schaffer}:
\bea
    E_M (1,+1)&=&\frac{A}{4}+P  \cr
    E_M (1,-1)&=&\frac{A}{4}-P  \cr
    E_M (1,0)&=&-\frac{A}{4}(1-2\sqrt{1+y^2})  \cr
    E_M (0,0)&=&-\frac{A}{4}(1+2\sqrt{1+y^2}),
\label{MuEnergy}
\eea
with
\bea
    P&=&\frac{1}{2}\mu_B B (g_{e^-}-g_{\mu^-}\frac{m_e}{m_{\mu}})
     \approx 5.76\times 10^{-9} B ({\rm eV/G})  \cr
    y&=&\frac{1}{A}\mu_B B (g_{e^-}+g_{\mu^-}\frac{m_e}{m_{\mu}})
     \approx 6.30\times 10^{-4} B ({\rm 1/G}).
\eea
The corresponding eigenstates are expressed in a ``natural" basis
$\vert S_{\mu}^z S_e^z>$ as:
\bea
   \vert 1,+1>_M &=& \vert ++>_M  \cr
   \vert 1,-1>_M &=& \vert -->_M \cr
   \vert 1,0>_M &=& c\  \vert -+>_M +\  s\ \vert +->_M \cr
   \vert 0,0>_M &=& -s\  \vert -+>_M +\  c\ \vert +->_M,
\label{MuState}
\eea
where $\vert +->_M$ means
$\vert S_{\mu}^z =\frac{1}{2},  S_e^z =-\frac{1}{2}>_M$, etc., and
\bea
    c&=&\frac{1}{\sqrt 2}[1+\frac{y}{\sqrt{1+y^2}}]^{1/2}  \cr
    s&=&\frac{1}{\sqrt 2}[1-\frac{y}{\sqrt{1+y^2}}]^{1/2}.
\eea
It is noted that the ($J=1, J_z=0$) state among $1S$ triplet and
$1S$ singlet state ($J=0, J_z=0$), which are both energy eigenstates
in the absence of external magnetic fields, mix with each other
in the presence of $\overrightarrow {B}$ and they are not energy eigenstates
any more. Thus it is understood that energy eigenstates $\vert 1,0>$ and
$\vert 0,0>$ are the states which approach to ($J=1, J_z=0$)
and ($J=0, J_z=0$) states, respectively,
when the magnetic field $\overrightarrow {B}$ vanishes.
However, ($J=1, J_z=\pm$) states among $1S$ triplet
remain as energy eigenstates even in the presence of $\overrightarrow {B}$.

Energy eigenvalues and the corresponding eigenstates for the antimuonium system
in the presence of external magnetic field $\overrightarrow {B}$
are obtained from Eqs.(\ref{MuEnergy})(\ref{MuState})
by interchanging $P\leftrightarrow -P$, $y\leftrightarrow -y$ and
$c\leftrightarrow s$. Thus the energy eigenvalues for the
antimuonium are
\bea
    E_{\overline{M}} (1,+1)&=&\frac{A}{4}-P  \cr
    E_{\overline{M}} (1,-1)&=&\frac{A}{4}+P  \cr
    E_{\overline{M}} (1,0)&=&-\frac{A}{4}(1-2\sqrt{1+y^2})  \cr
    E_{\overline{M}} (0,0)&=&-\frac{A}{4}(1+2\sqrt{1+y^2}),
\label{AntiMuEnergy}
\eea
and the corresponding eigenstates are
\bea
    \vert 1,+1>_{\overline{M}} &=& \vert ++>_{\overline{M}}  \cr
    \vert 1,-1>_{\overline{M}} &=& \vert -->_{\overline{M}} \cr
    \vert 1,0>_{\overline{M}} &=&
        s\  \vert -+>_{\overline{M}} + c\ \vert +->_{\overline{M}} \cr
    \vert 0,0>_{\overline{M}} &=&
          -c\  \vert -+>_{\overline{M}} + s\ \vert +->_{\overline{M}}.
\eea

Now we consider the $M-\overline{M}$ conversion in the presence of
static external magnetic fields. First we write down a useful formula
for the $M-\overline{M}$ conversion which was derived by
Feinberg and Weinberg a long time ago~\cite{Feinberg}. If there exists
an interaction ${\cal H}_{M\overline{M}}$ which would yield
a matrix element for conversion of $M$ into $\overline{M}$ equal to
\be
   <\overline M\vert {\cal H}_{M\overline{M}}\vert M>=\frac{\Delta }{2},
\ee
the mass matrix for the $M-\overline{M}$ system is written as
\be
   {\cal M}_{M\overline{M}}
              = \pmatrix{E_M&\frac{\Delta }{2} \cr
                         \frac{\Delta }{2}&E_{\overline{M}} \cr}.
\ee
Then the probability for a muonium atom of the state
$\vert M>$ to decay as antimuonium of the state $\vert \overline{M}>$ at all
is given by
\be
     P({\overline{M}})=\frac{\Delta ^2}
      {2 [\lambda ^2 + (E_M - E_{\overline{M}})^2 + \Delta ^2 ]},
\label{Probability}
\ee
where $\lambda =G_F^2 m_\mu ^5 /192\pi ^3$ is the muon decay rate and
$G_F$ is Fermi constant.

Before we study the dilepton contributions to the $M-\overline{M}$
conversion in the presence of static external magnetic fields,
we review the case when the effective Hamiltonian for
$M-\overline{M}$ transition is written in the $(V-A)\times (V-A)$ form
or $(V+A)\times (V+A)$ form~\cite{Schaffer}\cite{Jungmann},
\be
{\cal H}_{M\overline{M}}=
         \frac{G_{M\overline{M}}}{\sqrt 2}
      [\overline \mu \gamma_\lambda (1\mp \gamma_5)e]
                 [\overline \mu \gamma^\lambda (1\mp \gamma_5)e]+ H.c. ,
\label{Hold}
\ee
which arises in the left-right symmetric
models and the models with doubly-charged Higgs
bosons~\cite{Halprin}-\cite{Herczeg}.
In this case matrix elements for conversion of $M$
into $\overline{M}$ are given in a `` natural" basis
$\vert S_{\mu}^z S_e^z>$ as follows:
\bea
 _{\overline {M}}<++ \vert {\cal H}_{M\overline{M}}
         \vert ++>_M &=&
 _{\overline {M}}<-- \vert {\cal H}_{M\overline{M}}\vert -->_M \cr
   &=& _{\overline {M}}<+- \vert {\cal H}_{M\overline{M}}\vert +->_M \cr
   &=&  _{\overline {M}}<-+ \vert {\cal H}_{M\overline{M}}\vert -+>_M  \cr
    &=& \frac{\delta }{2}  \cr
  \cr
{\rm other \  elements} &=& 0,
\eea
with
\be
        \delta =\frac{16G_{M\overline{M}}}{\sqrt{2}\pi a^3},
\ee
where $a$ is the Bohr radius of the muonium $(m_r\alpha )^{-1}$ with
$m_r^{-1}=m_{\mu}^{-1}+m_e^{-1}$.
Thus we obtain,
\bea
   _{\overline {M}}<1,\pm 1 \vert {\cal H}_{M\overline{M}}
         \vert 1,\pm 1>_M &=& \frac{\delta }{2} \cr
 _{\overline {M}}<1,0 \vert {\cal H}_{M\overline{M}}\vert 1,0>_M
   &=& _{\overline {M}}<0,0 \vert {\cal H}_{M\overline{M}}\vert 0,0>_M \cr
   &=& c s \delta = \frac{\delta}{2 \sqrt{1+y^2}}.
\label{MMconv}
\eea
for the matrix elements in the ``energy eigenstate" representation.
Now it is straightforward from Eqs.(\ref{MuEnergy}), (\ref{AntiMuEnergy}),
(\ref{Probability}) and (\ref{MMconv}) to calculate the probability of a
muonium in the $\vert 1,\pm 1>$, $\vert 1,0>$ and
$\vert 0,0>$ states to decay as antimuonium.
The results are~\cite{Schaffer}\cite{Jungmann},
\be
     P^{(1,\pm1)}(\overline {M})=\frac{\delta^2}
         {2[\lambda^2 +4P^2+\delta^2]}
\ee
for the $\vert 1,+1>$ and $\vert 1,-1>$ states and
\bea
      P^{(1,0)}(\overline {M}) &=& P^{(0,0)}(\overline {M}) \cr
              &=&\frac{\delta^2}
         {2[(1+y^2)\lambda^2+\delta^2]}
\eea
for the $\vert 1,0>$ and $\vert 0,0>$ states.

It is noted that since the ($J=1, J_z=0$) and ($J=0, J_z=0$) states mix
with each other in the presence of external magnetic fields,
$M-\overline{M}$ conversions from $\vert 1,0>_M$ to
$\vert 0,0>_{\overline {M}}$ state and
from $\vert 0,0>_M$ to $\vert 1,0>_{\overline {M}}$ state are also possible.
Indeed, from the $M-\overline{M}$ transition matrix elements
\bea
 _{\overline {M}}<0,0 \vert {\cal H}_{M\overline{M}}
         \vert 1,0>_M &=& - _{\overline {M}}<1,0 \vert
                  {\cal H}_{M\overline{M}} \vert 0,0>_M  \cr
      &=&-\frac{y}{\sqrt{1+y^2}}\frac{\delta }{2},
\eea
we obtain
\bea
     P^{(1,0) \rightarrow (0,0)}(\overline {M}) &=&
      P^{(0,0) \rightarrow (1,0)}(\overline {M})    \cr
    &=& \frac{y^2 \delta^2}
         {2[(1+y^2)\lambda^2+(1+y^2)^2 A^2 + y^2 \delta^2]}
\eea
for the probability of a muonium of the $\vert 1,0>_M$ ($\vert 0,0>_M$) state
to decay as
antimuonium through the state $\vert 0,0>_{\overline {M}}$
($\vert 1,0>_{\overline {M}}$).
However these probabilities
are numerically extremely small and can be safely neglected
in the following discussion.

The assumption that each state is produced
with equal weight at the beginning gives
\be
  P^{\rm Tot}(\overline {M})= \frac{\delta^2}{4[\lambda^2 +4P^2+\delta^2]}
   + \frac{\delta^2}{4[(1+y^2)\lambda^2+\delta^2]},
\ee
for the ``total" propability of a muonium to decay as antimuonium.
The magnetic field dependence
of $P^{\rm Tot}(\overline {M})$ has been studied in
Refs.~\cite{Schaffer}\cite{Jungmann}. We plot the results for
dependence of $P^{\rm Tot}(\overline {M})$,
$\frac{1}{2}P^{(1,1)}(\overline {M})$, and
$\frac{1}{2}P^{(1,0)}(\overline {M})$ on $B$ in Fig.2.
Note that the probabilities are
normalized by $P^{\rm Tot}(\overline {M})\vert _{B=0}$ and
$G_{M\overline {M}}$ is taken to be $0.1 G_F$.

In the presence of static external magnetic fields, the degeneracy between
the $\vert 1,+1>_M$ and $\vert 1,+1>_{\overline {M}}$ states
(the $\vert 1,-1>_M$ and $\vert 1,-1>_{\overline {M}}$ states) breaks down
and the generated energy difference severely suppresses the conversion. In
fact, the
probability $ P^{(1,\pm1)}(\overline {M})$ becomes negligibly small when
$B$ is in the order of $10^{-1}$ G (see Fig.2-b).
On the other hand,
the $\vert 1,0>_M$ and $\vert 1,0>_{\overline {M}}$ states
(the $\vert 0,0>_M$ and $\vert 0,0>_{\overline {M}}$ states) remain degenerate
and thus the conversion persists up to the fields in the order of
$10^{3}$ G.  In the limit of large $B$, the $\vert 1,0>_M$
state becomes a pure $\vert -+>_M$ while
the $\vert 1,0>_{\overline{M}}$ state
becomes a pure $\vert +->_{\overline{M}}$, and thus the
matrix element
$_{\overline {M}}<1,0 \vert {\cal H}_{M\overline{M}}\vert 1,0>_M$
vanishes. Hence the probability
$P^{(1,0)}(\overline {M})$ reduces to zero in this
limit (see Fig.2-c below). By the same reason,
$P^{(0,0)}(\overline {M})$ vanishes in the large $B$ limit.
Finally we see from Fig.2-a that
in the case of the effective Hamiltonian being in the
$(V-A)\times (V-A)$ form or $(V+A)\times (V+A)$ form and
$G_{M\overline {M}}=0.1G_F$, the $M-\overline{M}$ conversion probability
is reduced to 50\% at a field strength as low as 0.26 G, to 35.8\% at
$B=1$ kG and to 1.2\% at $B=1$ T. The dependence of the
normalized probabilities on the coupling strength $G_{M\overline {M}}$
is negligibly small for $G_{M\overline {M}}< 1 G_F$.

Next we consider the $M-\overline{M}$ conversion in models with
dileptons. The gauge interaction of dileptons with leptons
is given by~\cite{Framb}
\bea
{\cal L}_{int}&=&
   -{g_{3l}\over2\sqrt2}X_\mu^{++}l^T C \gamma^\mu \gamma_5 l
   -{g_{3l}\over2\sqrt2}X_\mu^{--}\overline l\gamma^\mu \gamma_5 C\overline l^T
      \cr
 & &+{g_{3l}\over2\sqrt2}X_\mu^{+} l^T C
       \gamma^\mu (1-\gamma_5)\nu_l
    +{g_{3l}\over2\sqrt2}X_\mu^{-} \overline {\nu_l}
    \gamma^\mu (1-\gamma_5) C\overline l^T,
\label{Lint}
\eea
\noindent
where $l=e,\mu ,\tau $, and $C$ is the charge-conjugation matrix.
The gauge coupling constant $g_{3l}$ is
given approximately by $g_{3l}=1.19e$ for the SU(15) GUT model~\cite{Frama}
and by  $g_{3l}=g_2=2.07e$ for the
$SU(3)_L \times U(1)_X$ model~\cite{Framd},
where $e$ and $g_2$ are the electric charge and the $SU(2)_L$ gauge
coupling constant, respectively.
It is noted that the vector currents which couple to
doubly-charged dileptons $X^{\pm \pm}$ vanish due to Fermi statistics.
Through the doubly-charged-dilepton-exchange
diagram illustrated in Fig. 1, we obtain the following
effective Hamiltonian for the $M-\overline{M}$ conversion,
\be
{{\cal H}}_{M\overline{M}}^{Di}=
         \frac{G_{M\overline{M}}^{Di}}{\sqrt 2}
      [\overline \mu \gamma_\lambda (1-\gamma_5)e]
                 [\overline \mu \gamma^\lambda (1+\gamma_5)e]+ H.c.
\ee
where ${G_{M\overline{M}}^{Di}}/{\sqrt 2}=-g_{3l}^2/(8M_{X^{\pm \pm}}^2)$ and
$M_{X^{\pm \pm}}$ is the doubly-charged dilepton mass. This form
is obtained from Eq.(\ref{Lint}) and with help of the Fierz transformation.
It should be noted that the above effective Hamiltonian
is in the $(V-A)\times(V+A)$ form. The most stringent lower mass bound for
the doubly-charged dileptons at present is
$(M_{X^{\pm\pm}}/g_{3l})>340$ GeV (95\%C.L.)~\cite{Framb}. This gives
$G_{M\overline{M}}^{Di}<0.13G_F$.

With this effective Hamiltonian, we find that the
matrix elements for conversion of $M$ into $\overline{M}$ are given
in a `` natural" basis $\vert S_{\mu}^z S_e^z>$ as follows:
\bea
 _{\overline {M}}<++ \vert {\cal H}_{M\overline{M}}^{Di}
         \vert ++>_M &=&
 _{\overline {M}}<-- \vert {\cal H}_{M\overline{M}}^{Di}\vert -->_M
            =\frac{\hat{\delta }}{2} \cr
 _{\overline {M}}<+- \vert {\cal H}_{M\overline{M}}^{Di}\vert +->_M
   &=&  _{\overline {M}}<-+ \vert {\cal H}_{M\overline{M}}^{Di}\vert -+>_M
   = -\frac{\hat{\delta }}{2}  \cr
 _{\overline {M}}<+- \vert {\cal H}_{M\overline{M}}^{Di}\vert -+>_M
   &=&  _{\overline {M}}<-+ \vert {\cal H}_{M\overline{M}}^{Di}\vert +->_M
   = \hat{\delta } \cr
  \cr
{\rm other \  elements} &=& 0,
\eea
where
\be
        \hat{\delta} =-\frac{8G_{M\overline{M}}^{Di}}{\sqrt{2}\pi a^3}.
\ee
Since ${\cal H}_{M\overline{M}}^{Di}$ is in the
$(V-A)\times(V+A)$ form, the matrix elements
$_{\overline {M}}<++ \vert {\cal H}_{M\overline{M}}^{Di}\vert ++>_M$
and $_{\overline {M}}<+- \vert {\cal H}_{M\overline{M}}^{Di}\vert +->_M$
take different values, and
$_{\overline {M}}<+- \vert {\cal H}_{M\overline{M}}^{Di}\vert -+>_M$ and
$_{\overline {M}}<-+ \vert {\cal H}_{M\overline{M}}^{Di}\vert +->_M$ do
not vanish.

In terms of the ``energy eigenstates", the matrix elements for
$M-\overline{M}$ conversion are written as ,
\bea
   _{\overline {M}}<1,\pm 1 \vert {\cal H}_{M\overline{M}}^{Di}
         \vert 1,\pm 1>_M &=& \frac{\hat{\delta}}{2} \cr
 _{\overline {M}}<1,0 \vert {\cal H}_{M\overline{M}}^{Di}\vert 1,0>_M
   &=&(1-\frac{1}{2\sqrt{1+y^2}})\hat{\delta}  \cr
_{\overline {M}}<0,0 \vert {\cal H}_{M\overline{M}}^{Di}\vert 0,0>_M
   &=& -(1+\frac{1}{2\sqrt{1+y^2}})\hat{\delta}.
\eea
It is interesting to note that neither
$_{\overline {M}}<1,0 \vert {\cal H}_{M\overline{M}}^{Di}\vert 1,0>_M $ nor
$_{\overline {M}}<0,0 \vert {\cal H}_{M\overline{M}}^{Di}\vert 0,0>_M $
vanishes in the large $B$ (i.e., large $y$) limit.

Again using the formula (\ref{Probability}),
we obtain the following probabilities of a muonium to decay as antimuonium
in the models with dileptons:
\be
     P^{(1,\pm1)}_{Di}(\overline {M})=\frac{\hat{\delta}^2}
         {2[\lambda^2 +4P^2+\hat{\delta}^2]}
\ee
for the $\vert 1,\pm1>_M$ states,
\be
      P^{(1,0)}_{Di}(\overline {M}) =
       \frac{(2-\frac{1}{\sqrt{1+y^2}})^2\hat{\delta}^2}
         {2[\lambda^2+(2-\frac{1}{\sqrt{1+y^2}})^2 \hat{\delta}^2]}
\ee
for the $\vert 1,0>_M$ state and finally
\be
       P^{(0,0)}_{Di}(\overline {M}) =
      \frac{(2+\frac{1}{\sqrt{1+y^2}})^2\hat{\delta}^2}
         {2[\lambda^2+(2+\frac{1}{\sqrt{1+y^2}})^2 \hat{\delta}^2]}
\ee
for the $\vert 0,0>_M$ state.

As before we assume that each state is produced with equal weight at
the beginning, and we obtain,
\be
  P_{Di}^{\rm Tot}(\overline {M})=
 \frac{\hat{\delta}^2}{4[\lambda^2 +4P^2+\hat{\delta}^2]}
   + \frac{(2-\frac{1}{\sqrt{1+y^2}})^2\hat{\delta}^2}
         {8[\lambda^2+(2-\frac{1}{\sqrt{1+y^2}})^2 \hat{\delta}^2]}
      +\frac{(2+\frac{1}{\sqrt{1+y^2}})^2\hat{\delta}^2}
         {8[\lambda^2+(2+\frac{1}{\sqrt{1+y^2}})^2 \hat{\delta}^2]}.
\ee
for the ``total" probability of a muonium to decay as antimuonium.
In the limit of $B=0$, we have
\bea
   P_{Di}^{\rm Tot}(\overline {M})\vert _{B=0}
       &=&\frac{3\hat{\delta}^2}{8[\lambda^2 +\hat{\delta}^2]}
         +\frac{9\hat{\delta}^2}{8[\lambda^2 +9\hat{\delta}^2]}  \cr
  &\approx& \frac{3\hat{\delta}^2}{2\lambda^2 },
\eea
which is the result first obtained in Ref.~\cite{Fujii}.

In Fig.3 we plot the magnetic field dependence
of $P_{Di}^{\rm Tot}(\overline {M})$,
$\frac{1}{2}P_{Di}^{(1,1)}(\overline {M})$,
$\frac{1}{4}P_{Di}^{(1,0)}(\overline {M})$, and
$\frac{1}{4}P_{Di}^{(0,0)}(\overline {M})$. They are all normalized by
$ P_{Di}^{\rm Tot}(\overline {M})\vert _{B=0}$ and we take
$G_{M\overline{M}}^{Di}=0.1 G_F$.
As in the case of $P^{(1,\pm1)}(\overline {M})$,
the probability $P_{Di}^{(1,\pm1)}(\overline {M})$ becomes
negligibly small when $B$ reaches the order of $10^{-1}$G
since the magnetic field breaks the degeneracy of the
$\vert 1,+1>_M$ and $\vert 1,+1>_{\overline {M}}$ states (see Fig.3-b).
However, the $B$-dependences of $P_{Di}^{(1,0)}(\overline {M})$ and
$P_{Di}^{(0,0)}(\overline {M})$ are quite different from those of
$P^{(1,0)}(\overline {M})$ and $P^{(0,0)}(\overline {M})$ (see Fig.3-c,d).
Firstly, the $M-\overline {M}$ conversion through the channel
$\vert 0,0>_M \rightarrow \vert 0,0>_{\overline {M}}$ is much prefered.
Thus $P_{Di}^{(0,0)}(\overline {M})$ gives a dominant contribution to
$P_{Di}^{\rm Tot}(\overline {M})$.
Secondly, $P_{Di}^{(1,0)}(\overline {M})$ and $P_{Di}^{(0,0)}(\overline {M})$
remain finite in the large $B$ limit.
This is due to the fact that the matrix elements
$_{\overline {M}}<1,0 \vert {\cal H}_{M\overline{M}}^{Di}\vert 1,0>_M$ and
$_{\overline {M}}<0,0 \vert {\cal H}_{M\overline{M}}^{Di}\vert 0,0>_M$
do not vanish in the large $B$ limit when the effective Hamiltonian
is in the $(V-A)\times(V+A)$ form.
Interestingly enough,
$P_{Di}^{(1,0)}(\overline {M})$ starts to increase around $B=1$ kG
and partially compensates the decrease of $P_{Di}^{(0,0)}(\overline {M})$
in the region $B>1$ kG. Summing up each
contributions, we find that $P_{Di}^{\rm Tot}(\overline {M})$
is rather insensitive to the static external magnetic field.
In fact Fig.3-a shows that $P_{Di}^{\rm Tot}(\overline {M})$ is lowered to
$83\%$ in the region $0.2$ G$<B<300$ G and only to $67\%$ in the large
$B$ limit. At $B=1$ kG (1 T) the reduction is 22.4\% (32.9\%).
Again the dependence of the
normalized probabilities on the coupling strength $G^{Di}_{M\overline {M}}$
is negligibly small for $G^{Di}_{M\overline {M}}< 1 G_F$.

In conclusion, we have studied the magnetic field dependence of the
$M-\overline{M}$ conversion in the models with dileptons. We have found that
the conversion is rather insensitive to the strength of the magnetic fields.
If an experiment is performed in a magnetic field of 1 T and if a bound for
the conversion probability  $P(\overline{M})<10^{-10}$ is
gained~\cite{Jungmann}, then
a bound for the coupling strength, $G_{M\overline{M}}<1.8\times10^{-2}G_F$,
is obtained for the usual $(V\mp A)\times(V\mp A)$ type-Hamiltonian.
On the other hand, the models with dileptons give a more stringent bound
$G^{Di}_{M\overline{M}}<2.8\times10^{-3}G_F$.

\vspace{2cm}
\begin{center}
{\large\bf Acknowledgements}
\end{center}
\medskip
K.S. would like to thank
Professor G. zu Putlitz for the hospitality extended to him
when he visited Physikalisches Institut der Universit$\ddot{a}$t Heidelberg
in the summer of 1994 and for useful discussions.  We would like to thank
Professor K. Jungmann for introducing the work of
Refs.~\cite{Schaffer}\cite{Jungmann} to us,
which inspired us to start this work.

%------------------ REFERENCES --------------------------------------

%---------------------- Figure caption -------------------------
\newpage
\noindent
{\large\bf Figure caption}
\medskip

\noindent
Figure 1

\noindent
The doubly-charged dilepton exchange diagram for muon-antimuonium
conversion. The arrows show the flow of lepton number.

\medskip

\noindent
Figure 2

\noindent
The magnetic field dependence of the $M-\overline {M}$ conversion
probability with an effective $(V\mp A)\times (V\mp A)$ type-Hamiltonian:
(a) $P^{\rm Tot}(\overline {M})$; \ (b) $\frac{1}{2}P^{(1,1)}(\overline {M})$;
\ (c) $\frac{1}{2}P^{(1,0)}(\overline {M})$.
They are all normalized by
$ P^{\rm Tot}(\overline {M})\vert _{B=0}$ and $G_{M\overline {M}}=0.1 G_F$ is
assumed.

\medskip

\noindent
Figure 3

\noindent
The magnetic field dependence of the $M-\overline {M}$ conversion
probability in models with dileptons:
(a) $P_{Di}^{\rm Tot}(\overline {M})$;
\ (b) $\frac{1}{2}P_{Di}^{(1,1)}(\overline {M})$;
\ (c) $\frac{1}{4}P_{Di}^{(1,0)}(\overline {M})$;
\ (d) $\frac{1}{4}P_{Di}^{(0,0)}(\overline {M})$.
They are all normalized by
$ P_{Di}^{\rm Tot}(\overline {M})\vert _{B=0}$ and
$G_{M\overline {M}}^{Di}=0.1 G_F$ is assumed.

\end{document}